\documentclass{ws-ijmpa}

\def\beq{\begin{eqnarray}}    
\def\eeq{\end{eqnarray}}      

\def\al{\alpha}
\def\be{\beta}

\def\de{\delta}

\def\pa{\partial}

\begin{document}

\markboth{Bruno Gon\c{c}alves}
{Exact Foldy-Wouthuysen Transformation}

%
\catchline{}{}{}{}{}
%

\title{SOME ASPECTS OF THE EXACT FOLDY-WOUTHUYSEN 
TRANSFORMATION FOR A DIRAC FERMION}

\author{\footnotesize BRUNO GON\c{C}ALVES
}

\address{Departamento de F\'{\i}sica, ICE,
Universidade Federal de Juiz de Fora
\\
Juiz de Fora, CEP: 36036-330, MG,  Brazil
\\
brunoxgoncalves@yahoo.com.br}

\maketitle

\pub{Received (26 September 2008)}{}

\begin{abstract}
The Foldy-Wouthuysen transformation (FWT) is used to separate 
distinct components of relativistic spinor field, e.g. electron 
and positron. Usually, the FWT is perturbative, but in some 
cases there is an involution operator and the transformation 
can be done exactly. We consider some aspects of an exact FWT 
and show that, even if the theory does not admit an involution 
operator, one can use the technique of exact FWT to obtain the 
conventional perturbative result. Several particular cases 
can be elaborated as examples. 

\end{abstract}

\section{Introduction}

The Foldy-Wouthuysen transformation (FWT)\cite{BSO} is a useful method to 
extract physical information from Dirac equation in presence of external 
fields. For example, take the Dirac equation in external 
electromagnetic field and suppose that $\psi$ has the bi-spinor form
\beq
i\hbar\frac{\partial }{\partial t}\psi =
[c\overrightarrow{\alpha }\cdot(\overrightarrow{p}-\frac{e}{c}
\overrightarrow{A})+\beta mc^{2}+eA_0 ]\psi \,
\quad
\mbox{where}
\quad
\psi  = 
\left( \begin{array}{c}
\varphi  \\
\chi    \\
       \end{array}
\right)
e^{\frac{-imc^2t}{\hbar}}\;\;.
\label{expo}
\eeq

Substituting the second equation into the first one we get two 
equations that represent a mixture of the fields $\varphi$ and $\chi$. In 
order to separate the equations for $\varphi$ and $\chi$, one can perform the 
well known Foldy-Wouthuysen transformation\cite{Bjorken-Drell}. The first 
step is to define even and odd operators. An odd operator anticommutes 
with $\be$, it is mixing the two components of the wave function. Even 
operators commute with $\be$ and do not produce mixing.
 
In order to perform the FWT, the Hamiltonian is written in 
the form $H=\beta m +\varepsilon+\vartheta\,,$ where $\varepsilon$ 
are even operators and $\vartheta$ are odd ones. The usual perturbative 
FWT is an expansion in powers of (1/m). The transformed 
(purely even) Hamiltonian in the order $(1/m^3)$ has the form 
(see, e.g. Ref. \refcite{Bjorken-Drell})
\beq
H^{\prime\prime\prime} = 
\be \Big( m + \frac{\vartheta}{2m} + \frac{\vartheta^2}{8m^3} \Big) +
\varepsilon + \frac{1}{m^2}[\vartheta\,,[\vartheta\,,\varepsilon]] - 
\frac{1}{8m^2}[\vartheta\,,\dot\vartheta]\,.
\eeq

As it was already mentioned, this result is perturbative. On 
the other hand, in some cases there is the possibility to obtain 
an exact solution with separeted components of a Dirac field. 
In order to perform the EFWT, it is 
necessary to check whether the Hamiltonian of the theory 
falls into the class of models which admit the anticommuting 
involution operator $J = i\gamma_5\beta$. There are many papers 
describing the algorithm to perform this transformation, e.g. 
Refs.~\refcite{OT,diraceq,EK,nikitin}.

The operator $J$ is Hermitian, $J^\dagger = J$, and unitary, 
$JJ^\dagger = J^2 = 1$ at the same time. Moreover, it 
anticommutes both with the Hamiltonian 
$J\widehat{\cal H} + \widehat{\cal H}J = 0\,$ and with the 
$\beta$ matrix $J\beta + \beta J = 0$.
If this condition is satisfied the transformed Hamiltonian has the 
form
\beq
H^{tr}\,=\,UHU^{*}\,=\,\beta[\sqrt{H^2}]^{EVEN}
+ J[\sqrt{H^2}]^{ODD}\,.
\label{tr}
\eeq

As a result of EFWT one obtains $H^2$ and, afterwards, it is possible to 
expand the square root of the operator $H^2$ in some parameter 
considered small in the theory. For example, if one chooses the 
parameter $1/m$, the resulting Hamiltonian describes the theory 
in the non-relativistic limit. Let us note that the result of 
this procedure can, in principle differ from the one of 
the perturbative FWT \cite{diraceq}. Once the transformed 
Hamiltonian is obtained, in order to make some 
interpretations of the result, the next step is to present 
the Dirac fermion $\psi$ in the bi-spinor form (\ref{expo})
and use the equation $i\hbar\pa_t\psi=H\psi$ to derive the Hamiltonian 
for the field $\varphi$. 

\section{A method for the EFWT}
\label{EFWT}

If the theory has many external fields, the operator $H^2$ in 
equation (\ref{tr}) can becomes cumbersome.
In order to avoid this situation, one can use some notations which can be
useful in many different theories. Let us consider a Hamiltonian 
in general form 
\beq
H &=&\,\be mc^2 +\, \be q + \,\alpha^j K^i_j\partial_i+\,\alpha^ig_i\,,
\qquad
\mbox{where}
\qquad
K^i_j = -i \hbar c (\de^i_j + T^i_j v )\,.
\label{not}
\eeq

In these notations, the matrix $T^i_j$ has numerical 
parameters and q is a constant. The external fields are 
introduced by the scalar function $v$ and the vector $g_i$. 
According to the standard prescription \cite{EK}, the first 
step in deriving the EFWT is to calculate the square 
of the Hamiltonian $H^2$. A direct calculus gives, after 
certain algebra, the following result for the 
square of the Hamiltonian $H^2$
\beq
H^2 &=& m^2c^4\,+\,K^{il}K^m_l\pa_i\pa_m\,
+\,2\,K^i_jg^j\pa_i\,+\,g^2\,+\,
K^{il}\pa_i(K^m_l)\pa_m\,+\,K^i_j\pa_i(g^j)
\nonumber
\\
&+&
\,i\Sigma_k\epsilon^{jlk}K^i_j\pa_i(K^m_l)\pa_m
\,+\,
i \Sigma_k \epsilon^{jlk}K^i_j\pa_i(g_l) 
\,+\, 2mc^2 q \,+\, 
\alpha^j K^i_j\partial_i(q)\be \,.
\label{resultado}
\eeq

Let us notice that the  components of the matrix $K^i_j$ depend 
exclusively on $v$ while the components of the vector 
$g_i$ can also depends on other external fields.
The expression (\ref{resultado}) is rather general, but it can also be 
formulated in case that $T^i_j$ and $g_i$ depend on 
$\alpha$ matrices\cite{BSO}. Moreover, this result includes 
various other particular known cases. The main point is to 
choose the correct interaction term $g_i$. In order to 
illustrate this 
fact, if we put $v=0$, the matrix $K^i_j$ becomes 
$-i\,\hbar \,c \,\delta^i_j$. Then, with the term $g_i$ equal 
to zero, we have the free particle case. In case $g_i=-eA_i$, 
we have the particle is in presence of magnetic field and in case 
$g_i=-eA_i+\alpha_i \mu_I \eta 
\overrightarrow{\Sigma}.\overrightarrow{B}\,,$ 
we meet the particle with the anomalous magnetic moment in the  
magnetostatic field $\overrightarrow{B}$. All these cases were 
tested and gave the same well-known results from the literature 
\cite{EK}. 

\section {Semi-exact transformation}
\label {SET}

Let us now study the case when the Hamiltonian does not anticomute with 
the involution operator. The EFWT is not allowed anymore, but it is 
possible to obtain an even Hamiltonian using equation (\ref{tr}). 
The method can be explained using a simple example. Let us suppose 
a Dirac particle interacting with constant and uniform magnetic 
and scalar electromagnetic potential fields. The Hamiltonian reads
\beq
H\,=\,c\overrightarrow{\al}\cdot\overrightarrow{p}
- e\overrightarrow{\al}\cdot\overrightarrow{A}
+ e\Phi \,+ \be mc^2\,.
\label{H}
\eeq
Direct inspection shows that the term $e\Phi$ 
does not anticommute with 
the involution operator. 
Thus, the Hamiltonian (\ref{H}) does not enable one to perform 
the EFWT.

Let us make a modification of the term $e\Phi$, that is 
multiply it by the $\be$-matrix. The modified term 
anticommutes with the involution operator and now 
the exact transformation is perfectly 
possible. The main point is that, in the linear order in 
$1/m$, an extra factor of $\be$ has no effect. The reason is that, 
after deriving the final Hamiltonian operator, it will have the 
block diagonal structure. We are interested only in the upper 
block of Hamiltonian which is even (after transformation) to 
perform physical analysis. Therefore, it does not matter if this 
term is multiplied by $\be$ 
or not, because beta has a two block form with the unity matrix 
(in standard representation)
in the relevant block. As a result we arrive 
at what one can call semi-exact Foldy-Wouthuysen transformation,
because it is exact in only part of external fields and linear
in other external fields.
After all, the Hamiltonian we are going to deal with has the form
\beq
H\,=\,c\overrightarrow{\al}\cdot\overrightarrow{p}
- e\overrightarrow{\al}\cdot\overrightarrow{A}
+ e\be \Phi \,+ \be mc^2\,.
\label{H1}
\eeq
Following all the steps described in the introduction and using 
the result (\ref{resultado}) to the equation (\ref{H1}), the 
non-relativistic transformed Hamiltonian is
\beq
H^{tr} = \frac{1}{2m} 
\Big(\overrightarrow{p} - \frac{e}{c} \overrightarrow{A}\Big)^2 \,
+ e \Phi \,+ \frac{\hbar e}{2mc}\overrightarrow{B}\,.
\eeq
This result is in perfect agreement with the one obtained using the 
perturbative approach in Ref.~\refcite{BBS} if the torsion 
fields are equal to zero. The advantage here is that the result 
can be obtained in a much more economic way. The 
semi-exact Foldy-Wouthuysen transformation described above can also 
be applied to a more complex Hamiltonian such as Dirac particle 
interacting with constant magnetic and torsion field which 
were previously treated perturbatively in 
Refs.~\refcite{SHA,rytor}. The 
preliminary report on the result of the semi-exact calculation will be 
published in Ref.~\refcite{BSO2}.
\section{Conclusions}
In this paper, the relation between the two 
different methods used to obtain an even Dirac Hamiltonian 
were discussed. They were illustrated with simple examples, but can 
also be applied to more complex Hamiltonians.
In section (\ref{EFWT}), some notations were introduced and with them 
is was possible to write a very general solution that can be applied 
to many theories, for example Ref.~\refcite{BSO}. 
In section (\ref{SET}) there is a formulation, using an explicit 
example, of what was called the semi-exact Foldy Wouthuysen 
transformation, that is a 
transformation that makes the Hamiltonian even with a method 
of calculation very similar to the one used in the EFWT, for a 
Hamiltonian that does not admit the involution operator.

\section*{Acknowledgements}

I would like to thank Yuri N. Obukhov and Ilya L. Shapiro for useful 
discussions and advises in the course of this work. I am also grateful 
to FAPEMIG for support.


\end{document}